\documentclass[journal]{IEEEtran}
\usepackage[active]{srcltx}
\usepackage{amsfonts}
\usepackage{graphics}
\usepackage{eurosym}
\usepackage{amssymb}
\usepackage{pstricks}
\usepackage{shortvrb}
\usepackage{graphicx}
\usepackage{subfigure}
\usepackage{color}
\usepackage{multirow}
\usepackage{tabularx}
\usepackage{float}
\usepackage{ifpdf}
\usepackage{multicol}
\usepackage{graphicx}
\usepackage[cmex10]{amsmath}
\usepackage{amsthm}
\usepackage{float}
\usepackage{comment}
\usepackage{alphalph,etoolbox}
\usepackage{ushort}
\usepackage{mathabx}
\usepackage{booktabs}
\usepackage{algorithm}
\usepackage{algpseudocode}
\usepackage{tikz}
\usepackage{tikzpeople}
\usepackage{makecell}
\usepackage{hyperref}
\usepackage{textcomp}
\usepackage{todonotes}
\usepackage{pgfplots}
\usepackage[table]{xcolor}
\usepackage{colortbl} 
\usepackage{bm}
\usepackage{tikz}
\usetikzlibrary{shapes.geometric, arrows.meta, positioning}

\tikzstyle{process} = [rectangle, rounded corners, minimum width=3.5cm, minimum height=1cm, text centered, draw=black, fill=blue!10]
\tikzstyle{arrow} = [thick, ->, >=stealth]
\usetikzlibrary{arrows.meta, positioning, shapes.geometric}

\setuptodonotes{inline}

\patchcmd{\subequations}{\alph{equation}}{\alphalph{\value{equation}}}{}{}

\newcolumntype{?}[1]{!{\vrule width #1}}
\floatname{algorithm}{Algorithm}

\allowdisplaybreaks

\usepackage[style=ieee]{biblatex}
\AtBeginBibliography{\small}

\pgfplotsset{compat=1.18}

\title{Fast Analytical Method for Resource Adequacy Assessment in Renewable-Rich Power Systems}
\author{Seyed Amir Hosseini, Mohammad Reza Karimi Gharigh, Antonello Monti, Peter Palensky, Saeed Peyghami}

\begin{document}

\maketitle
	
\begin{abstract}
This paper proposes a novel analytical method for resource adequacy assessment in renewable-rich power systems. The approach begins by applying the stress–strength method to estimate two key resource adequacy indices: loss of load expectation (LOLE) and expected energy not supplied (EENS). This method requires defining the probability distribution functions (PDFs) for both the generation side and the load side. The load is assumed to follow a normal probability distribution, while the generation-side probability distribution is derived from the combined PDFs of conventional, wind, and solar power sources. The approach models the cumulative impact of uncertainties on resource adequacy using the central limit theorem (CLT), and improves accuracy with a cumulant-based Gram–Charlier expansion when the CLT alone is insufficient. The effectiveness of the proposed method is validated through Monte Carlo simulations conducted on the IEEE single-area and multi-area RTS-96 test systems. The results demonstrate that the proposed method achieves a substantial reduction in computational time, which is crucial in the context of operational planning, without sacrificing accuracy. Furthermore, its performance improves as renewable penetration increases, making it particularly suitable for future power systems operation and planning.

\end{abstract}
	
\begin{IEEEkeywords}
Power system reliability, resource adequacy, stress–strength method, Gram–Charlier series, Monte Carlo simulation.
\end{IEEEkeywords}

\vspace{-0.4cm}
\section{Introduction}
\label{sec:introduction}
In recent years, modern power systems have increasingly integrated renewable energy sources (RESs), such as solar parks and wind farms~\cite{merabet2024power}. These resources offer a more efficient and sustainable alternative to traditional power grids, promoting decentralization and reducing carbon emissions in line with the Paris Agreement targets~\cite{khan2021energy}. However, the intermittent and unpredictable nature of RESs continues to hinder their large-scale adoption and poses challenges to system reliability, increasing the complexity of maintaining a stable and reliable power system~\cite{peyghami2020overview}.

Reliability assessment in power systems is essential to ensure stable and safe grid operation by accounting for outages, failures, and the repair of infrastructure components. Power system reliability is commonly evaluated using three hierarchical levels (HLs). \mbox{HL-I} assesses generation adequacy, ensuring sufficient capacity to meet demand without considering transmission or distribution. \mbox{HL-II} includes the transmission network, examining how constraints and failures affect power delivery and system reliability. \mbox{HL-III} provides a comprehensive assessment by incorporating generation, transmission, and distribution, and evaluates customer-level reliability metrics such as outage frequency and duration. These HLs offer a systematic framework for assessing system performance and supporting reliable power system development~\cite{Singh2018, li2014risk}.

There are two primary methods to assess the reliability of the power systems, which are based on deterministic and probabilistic approaches~\cite{dong2010probabilistic,billinton1992reliability}. A deterministic method in reliability evaluation assumes fixed values for system parameters and examines specific failure scenarios without accounting for uncertainty or the likelihood of various outcomes. It generally focuses on worst-case or predefined conditions, with the N-1 criterion being an example meaning that the system must be able to withstand the failure of any single component without causing widespread outages or instability \cite{heylen2019multi}. A probabilistic method in reliability evaluation uses statistical models to account for uncertainty and variability in system components, considering the likelihood of different failure scenarios and their impact on system performance. Examples of this method include fault tree analysis (FTA), reliability block diagrams (RBD), event tree analysis (ETA), minimum cut set, Markov chains, and Monte Carlo simulation (MCS) \cite{hosseini2025exploring,billinton1970power}. In power system reliability assessment, the MCS method has several advantages over other probabilistic approaches. It can model complex interactions among system components and simulate uncertainties such as fluctuating demand and variable generation \cite{li1994reliability}. By considering multiple failure combinations, it provides a more comprehensive evaluation of system reliability than methods like FTA or RBD \cite{singh2022uncertainty}. It also produces probability distributions and confidence intervals that support better decision-making \cite{8506413}. However, as system size increases, computational complexity rises sharply, and simulation times can extend from hours to days even on high-performance platforms. In planning-level reliability studies, time is not a critical factor since these analyses are conducted offline to support long-term planning decisions. However, in operational planing—where the focus is on ensuring system stability, resource adequacy, and real-time dispatch—timeliness is essential, and fast computational methods are crucial for effective decision-making \cite{ding2014operational,yaghoubi2025comprehensive}. Therefore, fast and efficient analytical approaches are essential, as they provide timely insights and support rapid decision-making.

Analytical methods for evaluating the reliability of wind farms have been extensively explored in the literature. Among these methods, a more advanced approach is presented in \cite{dobakhshari2009reliability}, which develops a Markov model to represent the output power states of a wind farm by extending a single-turbine model to account for multiple turbines. It models a wind farm's output using a Markov approach and simplifies the system by clustering power states to reduce complexity. However, it suffers from key flaws, including arbitrary clustering without justification, lack of validation, and an unrealistic assumption that turbines operate independently despite being exposed to the same wind conditions—leading to implausible output scenarios that weaken the model’s accuracy. The last issue has been fixed in \cite{sulaeman2016wind}. This reference considers that the output power of wind turbines in a wind farm has correlation and proposes converting wind speed data into power output and then clustering the wind farm outputs into bands (states) using similarity (Euclidean distance). Each band represents a range of wind speeds with similar power output—e.g., all wind speeds producing 0 MW are grouped into one band. The banding approach oversimplifies wind power variability by grouping different wind speeds into fixed output states. This method treats the wind farm like a derated conventional generator, ignoring the inherently stochastic nature of wind energy. As a result, it reduces modeling accuracy. Several studies like \cite{metivier2020efficient,weng2025probability} have explored the use of polynomial chaos expansion (PCE) for uncertainty quantification in power system studies. However, as the number of renewable energy sources increases, PCE-based methods may still become inefficient due to the high dimensionality of uncertainties, the strong nonlinear and intermittent behavior of renewable generation, and potential correlations among different sites. In addition, accurately representing the system response often requires fine discretization and high-order expansions, which can significantly increase the computational cost in large-scale renewable-rich systems.

Despite prior efforts, existing methods rely on unrealistic or overly simplified assumptions. These limitations reduce modeling accuracy and fail to capture the true characteristics of generation resources. In this paper, we present an efficient framework for power system reliability assessment that integrates probabilistic modeling of both generation and load with the classical stress-strength approach. Our framework accounts for the discrete and stochastic nature of generation units, where uncertainties arise both from random unit outages (failures) and from variable renewable generation output. The core idea relies on the cumulative behavior of these uncertainties on resource adequacy rather than their individual characteristics. Therefore, central limit theorem is employed to model this cumulative behavior approximating with normal distribution. To enhance the accuracy of this approximation, we employ a cumulant-based method, the Gram–Charlier expansion, to correct the generation probability distribution, improving the accuracy of reliability measures when the central limit theorem alone is insufficient. The proposed analytical approach is validated against Monte Carlo simulations, demonstrating that it delivers accurate reliability estimates while significantly reducing computational time. This combination of accuracy, efficiency, and practical applicability distinguishes our work from existing approaches and enables reliable assessment of small- to large-scale power systems without the need for computationally intensive Monte Carlo simulations.

The rest of this article is organized as follows. Section \ref{sec:Reliability_indices_calculation} provides stress–strength analysis for evaluating power system reliability indices. Section \ref{sec:3} presents the PDFs for various types of power plants, while Section \ref{sec:4} focuses on the generation-side PDFs of the overall power system. Section \ref{sec:Simulation} details the simulation studies conducted, and finally, Section \ref{sec:Conclusion} summarized the paper outcomes.

\vspace{-0.3cm}
\section{Stress–Strength Analysis for Power System Reliability}
\label{sec:Reliability_indices_calculation}
\begin{figure}[b]
\centering
\includegraphics[width=0.89\columnwidth]{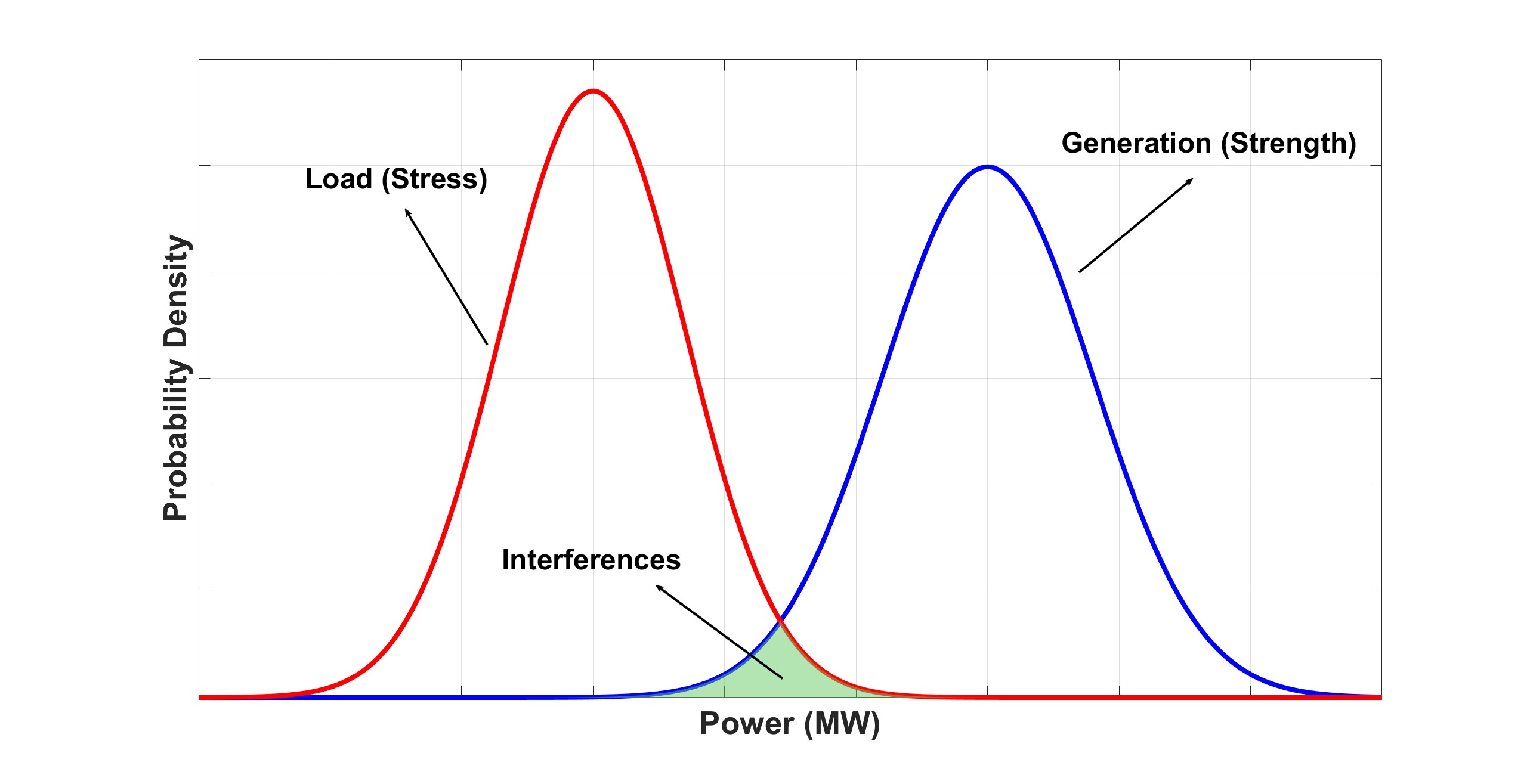}
\vspace{-0.1cm}
\caption{Stress-strength interference theory diagram}
\label{fig:stress1}
\vspace{-0.02cm}
\end{figure}

The stress-strength method is a reliability analysis approach used to determine the likelihood that a component or system will function properly without failing under applied conditions. In this paper, the stress-strength method is employed to calculate LOLE and EENS indices in power system. In this method, the system load is treated as the ``stress'' applied to the system, while the available generation capacity is treated as the ``strength'' as shown in Fig. \ref{fig:stress1}. Both are modeled as random variables with known or estimated PDFs, denoted by \( f_L(l) \) for load and \( f_G(p) \) for generation.

The probability that the load exceeds the generation, representing the likelihood of system failure, is given by:

\vspace{-0.25cm}
\begin{equation}\label{eq:lolp}
\text{LOLP} = \int_{0}^{\infty} \int_{0}^{l} f_G(g) f_L(l) \, dg \, dl
\end{equation}

The LOLE is then calculated by multiplying the LOLP by the number of time intervals, T (typically hours), in the assessment period:

\vspace{-0.25cm}
\begin{equation}\label{eq:lole}
\text{LOLE} = \text{LOLP} \times T
\end{equation}


The EENS quantifies the expected magnitude of unsupplied energy during periods of inadequacy and is given by:
\begin{equation}\label{eq:eens}
\text{EENS} = \int_{0}^{\infty} \int_{0}^{l} (l - g) f_G(g) f_L(l) \, dg \, dl
\end{equation}

To apply this method, the PDFs of both generation and load must be determined. The power system load can be modeled as a normal distribution \cite{billington1988power}. However, the PDF of generation, as described in Equation (\ref{eq:00}), includes the PDFs from conventional power plants (\( f_{conv} \)), wind turbines (\( f_{wt} \)), and solar panels (\( f_{sp} \)), requires further analysis. These generation PDFs will be derived and discussed in the following section.
\vspace{-0.01cm}
\begin{equation} \label{eq:00}
f_{G} = f_{conv} + f_{wt} + f_{sp}
\end{equation}

\vspace{-0.25cm}
\section{Probability Distribution Functions of Power Plants}
\label{sec:3}

The probability density function (PDF) characterizes the likelihood distribution of a continuous random variable, effectively capturing the inherent randomness and uncertainty in variables such as wind speed and solar radiation~\cite{gupta2020fundamentals}. Complementing the PDF, the cumulative distribution function (CDF) quantifies the probability that the variable assumes a value less than or equal to a given threshold. Mathematically, for a continuous random variable \( X \) with PDF, \( f(x) \), the CDF is defined as:
\vspace{-0.25cm}
\begin{equation}\label{eq:2}
F(X) = P(X \leq x) = \int_{0}^{x} f(x) \,dx
\end{equation}

This framework forms the basis for probabilistic modeling crucial to uncertainty analysis in renewable energy systems. The following subsections specify suitable PDFs for conventional power plants, wind farms, and solar parks.

\vspace{-0.05cm}
\subsection{PDF of Conventional Power Plants}
\label{subsec:PDF_of_Conventional_power_plants}
The PDF of a conventional power plant's output can be modeled by considering a generation unit with maximum power \( P_{\text{max}} \). The plant is available with probability \( A_c \) and unavailable with probability \( 1 - A_c \). Let \( P \) be the random variable representing the power output. Its PDF, \( f_{conv}(P) \), is defined as:
\begin{equation}\label{eq:01}
f_{conv}(P) = A_c \delta(P - P_{\max}) + (1 - A_c)\delta(P)
\end{equation}

\noindent where \(\delta(.)\) is the Dirac delta function.

This model represents a two-state, Bernoulli, generator that operates either at full capacity or is completely offline.

\subsection{PDF of Wind Turbine Output Power}
\label{subsec:PDF_of_Wind_Turbine_Output_Power}
A wind turbine operates only between the cut-in speed ($v_{c-in}$), when wind is strong enough to generate power, and the cut-off speed ($v_{c-out}$), when it shuts down to prevent damage from excessive wind, as illustrated in Fig. \ref{fig:wt}. The Weibull distribution is the most commonly used PDF in modeling wind velocity randomness \cite{PETKOVIC2014133}. The Weibull PDF for wind speed, $f_{w}(v)$, is given by:

\begin{figure}[b]
\centering
\includegraphics[width=0.75\columnwidth]{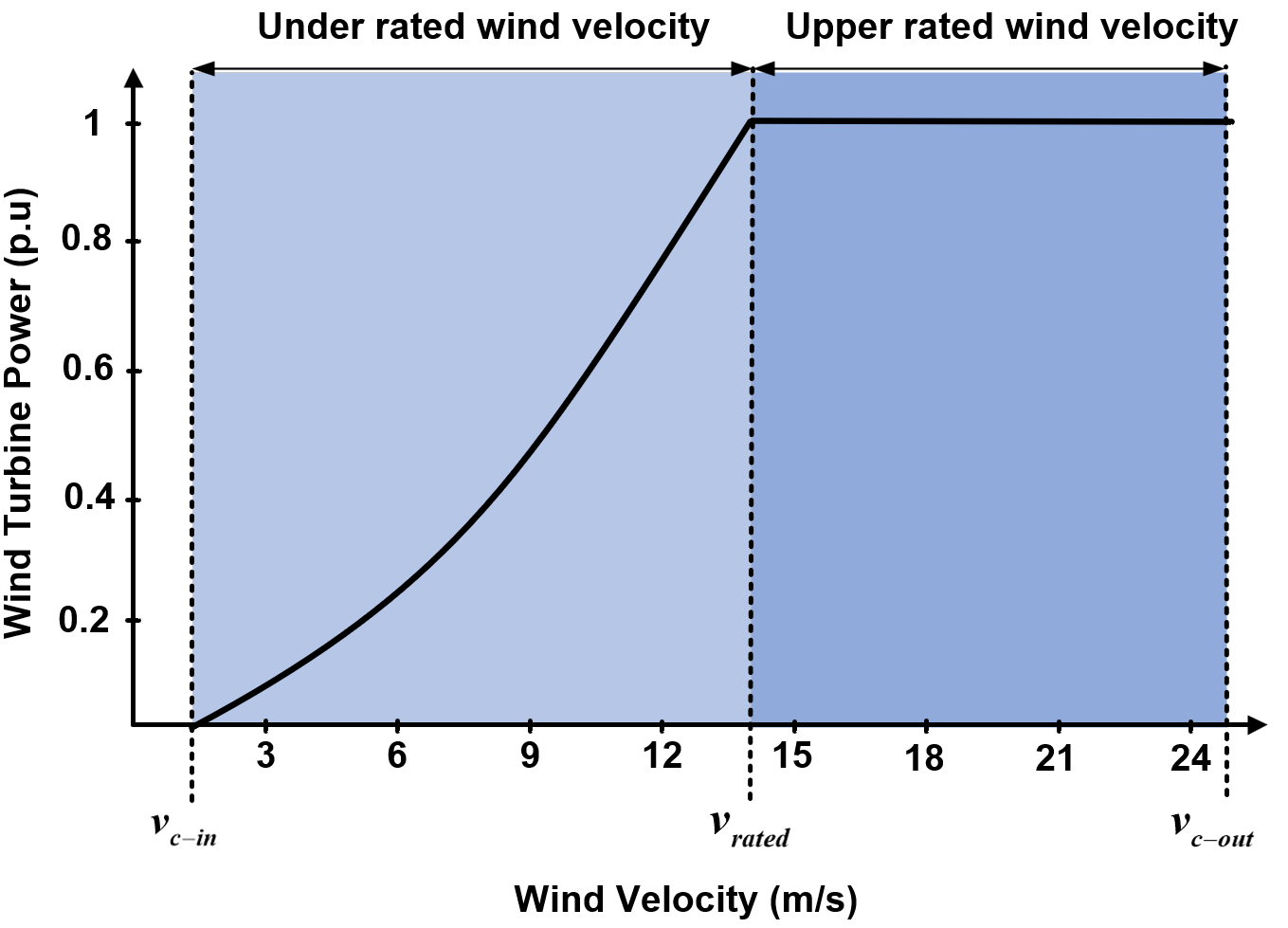}
\vspace{-0.1cm}
\caption{Wind turbine output power vs. wind speed.}
\label{fig:wt}
\vspace{-0.1cm}
\end{figure}
\begin{equation}\label{eq:1}
f_{w}(v) = \frac{k}{c} \left(\frac{v}{c}\right)^{k-1} e^{-\left(\frac{v}{c}\right)^k}
\end{equation}

\noindent where $v$, $k$, and $c$ are wind speed (m/s), shape parameter, and scale parameter, respectively.

The CDF of the wind speed, $F_w(v)$, is defined as:
\begin{equation}\label{eq:3}
F_w(v) = 1 - e^{-\left(\frac{v}{c}\right)^k}, \quad v \geq 0
\end{equation}

The power extracted from a wind turbine, \( P_{wt}\), can be calculated as below:  
\begin{equation}\label{eq:6}
P_{wt} = \frac{1}{2} \rho \pi R^2 C_p v^3 = H v^3
\end{equation}

\noindent where \( \rho \) is the air density (kg/m\(^3\)), \( R \) is the turbine blades radius (m), \( v \) is the wind velocity (m/s), and \( C_p \) is the power coefficient, which represents the efficiency of the turbine and depends on the tip speed ratio as well as the pitch angle in degrees. For simplicity, $\frac{1}{2} \rho \pi R^2 C_p$ is represented by $H$.

To find the PDF of the wind turbine output power from Equations (\ref{eq:1}) and (\ref{eq:6}), the following transformation formula is utilized \cite{brockhoff2018introduction}.
\begin{equation}\label{eq:7}
f_{wt}(P_{wt}) = f_w(v) \left| \frac{dv}{dP_{wt}} \right|
\end{equation}

Equation \eqref{eq:8} shows the wind velocity as a function of the wind turbine output power and the absolute value of the derivative of \( v \) with respect to \( P_{wt} \).
\begin{equation}\label{eq:8}
v = \left( \frac{P_{wt}}{H} \right)^{\frac{1}{3}}, \quad \left| \frac{dv}{dP_{wt}} \right| = \frac{1}{3}  H^{-\frac{1}{3}}{P_{wt}}^{-\frac{2}{3}}
\end{equation}

Based on Equations~\eqref{eq:7} and~\eqref{eq:8}, the PDF of the wind turbine output power, \( P_{wt} \), denoted as \( f_{\text{wt}}(P_{wt}) \) in Equation~\eqref{eq:9}, is defined piecewise to reflect the different operational states of the wind turbine.
\begin{equation}\label{eq:9}
\resizebox{\columnwidth}{!}{$
f_{wt}(P_{wt}) =
\begin{cases}
\delta_0 \cdot \delta(P_{wt}) &  P_{wt} = 0 \\[10pt]
\displaystyle
\frac{k}{3 \, c^k \, H^{k/3}} \cdot P_{wt}^{\frac{k}{3} - 1} \cdot 
e^{- \left( \frac{P_{wt}}{H c^3} \right)^{k/3}}, &  P_{\text{min}} \leq P_{wt} < P_{\text{max}} \\[10pt]
\delta_{\max} \cdot \delta(P_{wt} - P_{\text{max}}) &  P_{wt} = P_{\text{max}} \\[10pt]
0 & \text{otherwise}
\end{cases}
$}
\end{equation}

In this equation, when wind speeds fall outside the turbine’s operational range—specifically, below the cut-in speed or above the cut-out speed—the output power drops to zero (\( P_{wt} = 0 \)). This behavior is captured using a Dirac delta function \( \delta(.) \), scaled by a coefficient \( \delta_0 \), which represents the total probability mass associated with the zero-power state.
{\fontsize{10}{12}\selectfont
\begin{equation}\label{eq:delta0}
\delta_0 = F_w(v_{c-in}) + 1 - F_w(v_{c-off}) 
= 1 - e^{ -\left( \frac{v_{\text{c-in}}}{c} \right)^k } 
+ e^{ -\left( \frac{v_{\text{c-off}}}{c} \right)^k }
\end{equation}
}

In Equation \eqref{eq:delta0}, \( F_w \), denotes the CDF corresponding to the wind speed PDF as described in Equation \eqref{eq:3}.

For power values between \( P_{\text{min}} = H v_{\text{c-in}}^3 \) and \( P_{\text{max}} = H v_{\text{rated}}^3 \), corresponding to wind speeds within the turbine’s operational range (from cut-in to rated speeds), the PDF takes the form of a modified Weibull distribution. This continuous part of the function characterizes the statistical distribution of power output under normal operating conditions.

At the rated power output, \( P_{wt} = P_{\text{max}} \), the function includes another Dirac delta component, \( \delta(P_{wt} - P_{\text{max}}) \), scaled by a constant \( \delta_{\max} \). This term captures the non-negligible probability of the turbine operating steadily at its rated output, which occurs when the wind speed is between the rated speed ($v_{c-rated}$) and the cut-out speed ($v_{c-out}$).

\vspace{-0.25cm}
{\fontsize{10}{12}\selectfont
\begin{equation}\label{eq:deltamax}
\delta_{\max} = F_w(v_{c-off}) - F_w(v_{\text{rated}}) 
= e^{ -\left( \frac{v_{\text{rated}}}{c} \right)^k } 
- e^{ -\left( \frac{v_{c-off}}{c} \right)^k }
\end{equation}
}

For all other values of \( P_{wt} \), the PDF is zero, reflecting the physical constraints on power generation.

Wind turbines are not always operational, as they experience failures and require repairs. By considering availability of the wind turbine,  \( A_{wt} \), the PDF for the wind turbine's output power can be expressed as follows:

\vspace{-0.2cm}
\begin{equation}\label{eq:12}
\resizebox{\columnwidth}{!}{$
f_{wt}(P_{wt}) =
\begin{cases}
(1 - A_{wt}) + \left( \delta_0 \cdot \delta(P_{wt}) \right), & P_{wt} = 0 \\[12pt]
A_{wt} \cdot \frac{k}{3 c^k H^{k/3}} \cdot P_{wt}^{\frac{k}{3} - 1} \cdot 
e^{ - \left( \frac{P_{wt}}{H c^3} \right)^{k/3} }, & P_{\min} \leq P_{wt} \leq P_{\max} \\[12pt]
A_{wt} \left( \delta_{\max} \cdot \delta(P_{wt} - P_{\max}) \right), & P_{wt} = P_{\max} \\[12pt]

0, & \text{otherwise}
\end{cases}
$}
\end{equation}

The value of \( f_{wt}(P_{wt}) \) at \( P_{wt}=0 \), is composed of two separate terms: the probability that the turbine is unavailable, represented by (\( 1 - A_{wt} \)), and the probability that it is available but producing zero power, given by \( \delta_0 \cdot \delta(P_{wt}) \), which occurs when the wind speed is below the cut-in speed or above the cut-out speed. \( \delta_0 \) is defined in Equation (\ref{eq:delta0}).

For power values within the operational range of the turbine, i.e., between \( P_{\min} \) and \( P_{\max} \), the distribution is scaled by the availability factor \( A_{wt} \), reflecting the fact that this power output is only possible when the turbine is available. 

At the rated power \( P_{wt} = P_{\max} \), the function again includes a Dirac delta component to model the nonzero probability of the turbine operating at its rated capacity. This term is weighted by both \( A_{wt} \) and \( \delta_{\max} \), accounting for the fact that this state occurs only when the turbine is operational and the wind speed lies between the rated speed and the cut-out speed. \( \delta_{\max} \) was defined in Equation (\ref{eq:deltamax}).

Finally, for all other values of \( P_{wt} \), \( f_{wt}(P_{wt}) \), is zero, as such values fall outside the physical operating range of the turbine.

\vspace{-.4cm}
\subsection{PDF of Solar Panel Output Power}
\label{subsec:PDF_of_solar_panel_Output_Power}

The Gamma distribution can be used to model solar radiation as it effectively captures the positive, skewed nature of the data \cite{YAN20213023}. Solar radiation values, which are always non-negative, often exhibit variability and skewness due to factors such as cloud cover and time of day \cite{augbulut2021prediction}. The Gamma distribution, with its shape parameter ($\alpha$) controlling the skewness and scale parameter ($\beta$) determining the spread, is a suitable choice for describing this variability \cite{johnson2019statistics}. By adjusting the parameters, the Gamma PDF can accurately represent the distribution of solar radiation, making it a widely adopted model for solar energy studies and related applications.

The PDF of the Gamma distribution is given by:

\begin{equation}\label{eq:13}
f(G) = \frac{G^{\alpha - 1} e^{-G/\beta}}{\Gamma(\alpha) \beta^{\alpha}}, \quad G > 0
\end{equation}

\noindent where \(G\) is the solar radiation on the solar panel in watts per square meter, (W/m²).

The output power of solar panels can be calculated based on solar radiation using the following equation:

\vspace{-0.2cm}
\begin{equation}\label{eq:14}
P_{sp} = S \cdot \eta  \cdot G
\end{equation}

\noindent where \(\ P_{sp}\) is the electric output power of the solar panel in watts (W), \(S\) is the area of the solar panel in square meters ($m^2$), and \(\eta\) is the efficiency of the solar panel.

When the sun is shining, solar panels generate electricity. Using Equations \eqref{eq:14}, \eqref{eq:88}, and \eqref{eq:89}, the PDF for the solar panel output power, $f_{sp}(P_{sp})$, is presented as follows:

\begin{equation}\label{eq:88}
f_{sp}(P_{sp}) = f(G) \left| \frac{dG}{dP_{sp}} \right|
\end{equation}

\begin{equation}\label{eq:89}
G =  \frac{P_{sp}}{S \eta} , \quad \left| \frac{dG}{dP_{sp}} \right| = \frac{1}{S \eta} 
\end{equation}

\begin{equation}\label{eq:115}
f_{sp}(P_{sp}) =\frac{P_{sp}^{\alpha - 1} e^{-P_{sp}/(S \eta\beta)}}{\Gamma(\alpha) (S\eta \beta)^{\alpha}}, \quad P_{sp} > 0
\end{equation}

In this paper, the solar panel output is modeled under the assumption that panels do not consistently operate at their maximum rated capacity. This assumption reflects real-world conditions in which irradiance levels, temperature variations, and degradation limit optimal performance. To accurately model solar panel performance, both the operational availability of the panels and the availability of sunlight during the day must be considered. In this context, power is produced only when sunlight is present and the panels are functioning properly. If it is nighttime or the panel is unavailable, it is assumed that no electricity is generated and the output power is set to zero. Based on equation \eqref{eq:115}, and accounting for the combined availability of solar panels and sunlight, \( A_{sp} \), the updated PDF of the solar panel output power is expressed as:

\vspace{-.2cm}
\begin{equation}\label{eq:16}
f_{sp}(P_{sp}) =
\begin{cases}
A_{sp} \cdot\left(\frac{P_{sp}^{\alpha - 1} e^{-P_{sp}/S \eta\beta}}{\Gamma(\alpha) (S \eta \beta)^{\alpha}} \right), &  P_{sp}\neq
0 \\
(1-A_{sp}) \cdot \delta (P_{sp}),  & P_{sp} = 0
\end{cases}
\end{equation}

where \(\delta(P_{sp})\) is the Dirac delta function.

\section{Probability Distribution Function of Power System Generation Side}
\label{sec:4}
Modern power systems comprise a mix of conventional power plants, wind farms, and solar parks, each characterized by its own PDF for their output power. In resource adequacy studies, HL-I, the total system generation is the sum of the outputs from these sources. The following subsections describe the proposed approach to accurately derive the PDF of total generation adequacy.

\vspace{-0.25cm}
\subsection{Central Limit Theorem (CLT)}
\label{ssec:Central_Limit_Theorem}

In probability theory, the central limit theorem (CLT) states that as the sample size increases, the distribution of the standardized sample mean approaches a normal distribution, regardless of the original population distribution, provided the population has a finite variance \cite{kwak2017central}. More precisely, for sufficiently large sample sizes, the sample mean \(\bar{X}\) is approximately normally distributed with mean equal to the population mean \(\mu\) and variance equal to the population variance \(\sigma^2\) divided by the sample size \(n\).

In modern power systems, the CLT suggests that the overall power generation distribution will approximate a normal distribution. It should be noted that since the number of wind turbines and solar panels is finite and they are subject to failures, the resulting PDF is not a perfectly pure normal distribution and has skewness and kurtosis. To evaluate deviations from a pure normal distribution in the overall power generation PDF, it is essential to first introduce fundamental concepts such as statistical moments and cumulants.

\vspace{-0.3cm}
\subsection{Moments}
\label{ssec:Moments}
In probability and statistics, moments are numerical measures that describe the shape of a function, particularly a probability distribution. Moments provide insights into fundamental characteristics such as central tendency and dispersion. Moments are often expressed as raw moments, which serve as foundational tools in statistical analysis \cite{grimmett2020probability}. The $n^{\text{th}}$ raw moment of a random variable \(x\) is given by:

\vspace{-.2cm}
\begin{equation}\label{eq:15}
\mu_{n}' = E[x^n] = \int_{-\infty}^{\infty} x^n f(x) \, dx
\end{equation}

\noindent where \( E[x^n] \) is the expected value of \( x^n \) and \( f(x) \) is the PDF of the random variable \( x \). Raw moments describe the overall shape of the distribution rather than deviations from the mean. The first raw moment is the mean, \(\mu\), while higher-order raw moments provide additional shape-related information.

\vspace{-0.3cm}
\subsection{Cumulants}
\label{ssec:Cumulants}
A cumulant is a statistical measure that describes the shape of a probability distribution, similar to moments but with properties that make them more useful in certain contexts, such as probability theory and statistical physics. Cumulants provide a way to characterize a distribution in terms of its deviations from a normal distribution \cite{kendall1948advanced}.

Cumulants, \(\kappa_n\), are related to moments, but they are defined through the cumulant-generating function (CGF), which is the logarithm of the moment-generating function (MGF). The CGF is given by \cite{pho2019moment}:

\begin{equation}\label{eq:17}
K(t) = \log E[e^{tX}]
\end{equation}

The $n^{th}$ cumulant (\(\kappa_n\)) is obtained by taking the \textit{\(n^{th}\) derivative} of \( K(t) \) and evaluating it at \( t = 0 \) \cite{lukacs1970characteristic}:

\begin{equation}\label{eq:18}
\kappa_n = \left. \frac{d^n K(t)}{dt^n} \right|_{t=0}
\end{equation}



The cumulants can be computed from the raw moments using the recursive relation shown in Equation ~\eqref{eq:cum}. 

\vspace{-.2cm}
\begin{equation}\label{eq:cum}
\kappa_n = \mu'_n - \sum_{m=1}^{n-1} \binom{n-1}{m-1} \kappa_m \cdot \mu'_{n-m}
\end{equation}

The first few cumulants describe key characteristics of a probability distribution. The first cumulant, \(\kappa_1\), represents the mean, \(\mu\). The second cumulant, \(\kappa_2\), corresponds to the variance, \(\sigma^2\). The third cumulant, \(\kappa_3\), is associated with skewness and the fourth cumulant, \(\kappa_4\), relates to kurtosis, providing insight into the heaviness of the distribution's tails compared to a normal distribution.

Cumulants offer several advantages over moments in statistical analysis. One basic property about the cumulants is if \( x_1, x_2, \dots, x_m \) are independent random variables, then the cumulants of their sum exhibit an important property known as additivity. Specifically, the cumulants of the sum of independent random variables are the sum of the cumulants of each individual random variable. Mathematically, this is expressed as:

\vspace{-0.5cm}
\begin{equation}\label{eq:20}
\kappa_n(x_1 + \cdots + x_m) = \kappa_n(x_1) + \cdots + \kappa_n(x_m),
\end{equation}

This additivity makes it easier to analyze large sums or mixtures of random variables, since it allows us to work with each part separately instead of dealing with the whole at once.

\vspace{-0.3cm}
\subsection{Gram-Charlier Series}
\label{ssec:Gram-Charlier series}
The Gram–Charlier series is a mathematical expansion used to approximate probability distributions that deviate from normality \cite{chateau2017gram}. It represents the characteristic function, i.e., the Fourier transform of a PDF \cite{kendall1946advanced}. Similar to the Edgeworth expansion, it modifies the normal distribution by incorporating skewness and kurtosis through cumulants \cite{jondeau2001gram}. This approach can model power generation PDFs as deviations from a normal distribution.

First, the raw moments are calculated for each generation type as described in Equation \eqref{eq:15}, then converted into cumulants using Equation \eqref{eq:cum}, which capture deviations from normality. These cumulants are then used in a series expansion with Hermite polynomials to refine the normal distribution for a better fit. As a result, Equation~\eqref{eq:gram} is obtained that accurately represents the real-world distribution while accounting for distortions.

\vspace{-0.2cm}
{\fontsize{8}{9}\selectfont
\begin{align} \label{eq:gram}
f_{ps}(P) =\ & \frac{1}{\sigma_{ps} \sqrt{2\pi}} 
\exp\left(-\frac{(P - \mu_{ps})^2}{2 \sigma_{ps}^2}\right) \times \\
& \left[ 1 + \sum_{n=3}^{\infty} \frac{B_n\left(0,0,\kappa_{ps_3},\kappa_{ps_4},\ldots,\kappa_{ps_n}\right)}{n!} 
H_n\left(\frac{P - \mu_{ps}}{\sigma_{ps}}\right) \right] \nonumber
\end{align}
}

In this equation, \( f_{ps}(P) \) is the PDF of power system generation side. The first term represents the normal distribution, where \( \mu_{ps} \) and \( \sigma_{ps} \) denote the mean and standard deviation of power system generation, as defined in Equation \eqref{eq:mu}.

\vspace{-0.2cm}
\begin{equation}\label{eq:mu}
\begin{aligned}
\mu_{ps} &= \mu_1 + \mu_2 + \cdots + \mu_n \\
\sigma_{ps}^2 &= \sigma_1^2 + \sigma_2^2 + \cdots + \sigma_n^2
\end{aligned}
\end{equation}

The bracketed summation introduces corrections to the normal approximation by incorporating higher-order cumulants \( \kappa_{ps_3}, \kappa_{ps_4}, \ldots, \kappa_{ps_n} \), as defined in Equation~\eqref{eq:20}. These cumulants originate from various generation sources and account for non-normal characteristics of the distribution. The correction terms are expressed using Bell polynomials \( B_n(\cdot) \) and Hermite polynomials \( H_n(\cdot) \), which systematically refine the approximation beyond the second-order statistics.


Bell polynomials are used in probability theory to relate moments and cumulants. They provide a compact way to compute raw moments from cumulants and vice versa.

The complete Bell polynomial of order \( n \), denoted by \( B_n \), is defined as:

\vspace{-0.35cm}
\begin{equation}\label{complete_bell}
B_n(x_1, x_2, \ldots, x_n) = n!
\sum_{\substack{1j_1 + 2 j_2 + \cdots + n j_n = n}} 
 \prod_{i=1}^n \frac{(x_i)^{j_i} }{(i!)^{j_i} j_i!}
\end{equation}

In this equation, the parameter \( n \) denotes the total order or degree of the polynomial. It represents the size of the set being partitioned or the total “weight” of the partitions considered in the summation. The index \( i \) runs from 1 to \( n \) and corresponds to the size of each part in the partition. Each \( j_i \) is a non-negative integer representing the number of parts of size \( i \) included in a particular partition. The sequence \( (j_1, j_2, \ldots, j_n) \) collectively describes how the integer \( n \) is partitioned into smaller parts. The variables \( x_i \) are parameters associated with parts of size \( i \). Finally, the summation runs over all sequences \( (j_1, j_2, \ldots, j_n) \) satisfying the constraint $1 j_1 + 2 j_2 + \cdots + n j_n = n $.

Furthermore, the probabilist's Hermite polynomials \( H_n(x) \) are defined by the Rodrigues formula:

\begin{equation}\label{hermit}
H_n(x) = (-1)^n e^{\frac{x^2}{2}} \frac{d^n}{dx^n} \left( e^{-\frac{x^2}{2}} \right)
\end{equation}

The Bell and Hermit terms capture skewness, kurtosis, and other deviations from normality, allowing the PDF to accurately reflect non-normal characteristics such as asymmetry and heavy tails. Notably, the summation starts at \( n=3 \) in Equation \eqref{eq:gram}, because the first two moments are already captured by the normal distribution.

\begin{algorithm}[t]
    \caption{Proposed Fast Analytical Algorithm for HL-I }
    \label{alg:proposed_algorithm}
    \begin{algorithmic}[1]
        \State Calculate the PDF of each power source using \eqref{eq:01}, \eqref{eq:12}, or \eqref{eq:16}, based on its type. \label{alg_line:pdf}
        \State Compute raw moments and cumulants for each power source using \eqref{eq:15} and \eqref{eq:cum}. \label{alg_line:moments}
        \State Compute the aggregated cumulants, \( \mu \), and \( \sigma \) of total generation using \eqref{eq:mu} and \eqref{eq:20}. \label{alg_line:agg}
        \State Derive the PDF of total generation using the Gram-Charlier expansion formulated in \eqref{eq:gram}. \label{alg_line:gram}
        \State Convolve the obtained PDF of total generation with the load using the stress-strength method, and obtain LOLP, LOLE, and EENS using \eqref{eq:lolp}--\eqref{eq:eens}. \label{alg_line:convolve_reliability}
        \State Validate the obtained reliability indices via Monte Carlo simulation. \label{alg_line:validate}
    \end{algorithmic} 
\end{algorithm}

The proposed fast analytical method for HL-I reliability assessment is demonstrated in Algorithm 1. First, the probabilistic behavior of each generation source (conventional, wind, and solar) is modeled by deriving its PDF and corresponding statistical moments and cumulants. These parameters are then aggregated to obtain the mean, variance, and higher-order cumulants of the total system generation. Based on these aggregated statistics, the overall generation distribution is approximated using the Gram–Charlier expansion. Finally, the resulting generation PDF is combined with the load model through the stress–strength method to compute reliability indices, including LOLP, LOLE, and EENS, and the results are validated using Monte Carlo simulation.


\vspace{-0.02cm}
\section{Simulation}
\label{sec:Simulation}
To demonstrate the effectiveness of the proposed methods in Algorithm \ref{alg:proposed_algorithm}, they are applied to the IEEE single-area RTS-96 as well as on its three-area and five-area extensions. Detailed electrical reliability parameters for each unit are adopted from \cite{li2013reliability}.

\vspace{-0.02cm}
\subsection{Validation of Proposed PDFs for conventional, wind and solar power plants}
\label{ssec:s1}
This subsection validates the proposed PDFs for conventional, wind, and solar power plant outputs using Monte Carlo simulation. 

\begin{itemize}

\item \textbf{Conventional power plant}: The PDF of the output power is commonly modeled by a Bernoulli distribution. This means the unit can produce up to its maximum power \( P_{\text{max}} \) with a probability equal to its availability, and it is off, zero output, with the probability of unavailability, as described in Equation~\eqref{eq:01}. For example, consider a 50 MW conventional unit with an availability of 0.95 and an unavailability of 0.05. The PDF of this unit is illustrated in Fig.~\ref{fig:con} by two delta functions, the green delta represents the 50 MW output with probability of 0.95, while the red delta corresponds to zero output with probability of 0.05. 

To validate this model, Monte Carlo simulation was performed, and its results are shown as two gray bars in Fig.~\ref{fig:con}. The comparison clearly demonstrates strong agreement between the analytical Bernoulli model and the simulation outcomes.

\begin{figure}
\centering
\includegraphics[width=0.99\columnwidth]{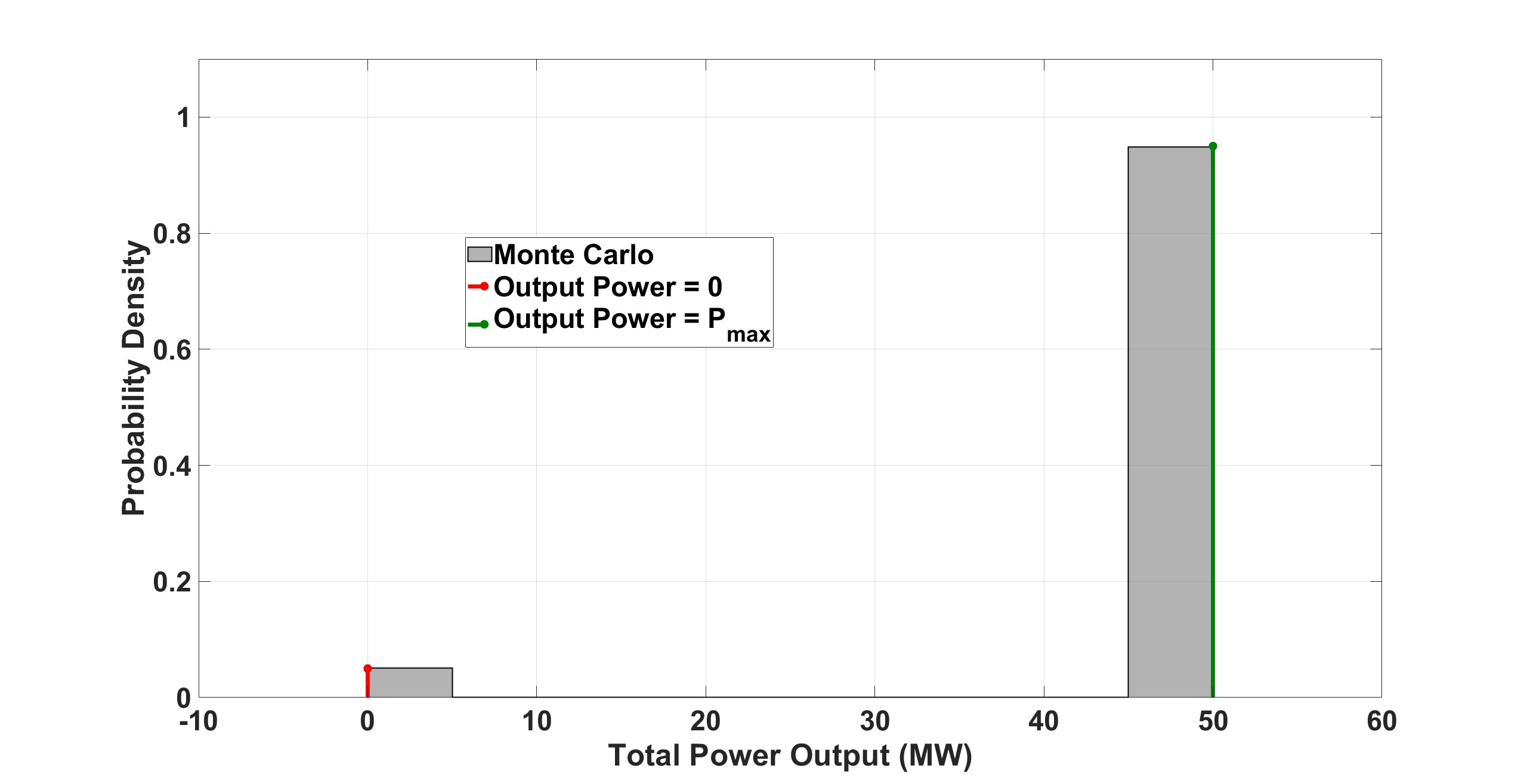}
\vspace{-0.05cm}
\caption{Analytical vs. Monte Carlo simulation of conventional power PDF}
\vspace{-0.1cm}
\label{fig:con}
\end{figure}

\item \textbf{Wind power plant}: It is assumed that the characteristics of a single wind turbine are as presented in Table \ref{tab:wind_params}. A hypothetical wind farm consisting of 40 identical turbines is considered, where all turbines are subject to the same wind conditions. The availability of wind turbines is assumed to be 0.95. The wind speed is assumed to follow a Weibull distribution with a shape parameter of \(k = 2.5\) and a scale parameter of \(c = 9.0\), North sea wind speed data obtained from \cite{nasa}.

\begin{table}
\centering
\vspace{-0.01cm}
\caption{Wind Turbine Parameters}
\label{tab:wind_params}
\begin{tabular}{llc}
\hline
\textbf{Parameter} & \textbf{Description} & \textbf{Value} \\
\hline
$P_{rated}$ & Rated Capacity (MW) & 2.4 \\
$v_{\text{c-in}}$ & Cut-in wind speed (m/s) & 4 \\
$v_{\text{rated}}$ & Rated wind speed (m/s) & 12.5 \\
$v_{\text{c-off}}$ & Cut-off wind speed (m/s) & 25 \\
$\rho$ & Air density (kg/m\textsuperscript{3}) & 1.225 \\
$R$ & Blade radius (m) & 40 \\
$C_p$ & Power coefficient & 0.4 \\
\hline
\end{tabular}
\end{table}

Using the analytical PDF derived in Equation~\eqref{eq:12}, the equivalent power output PDF for the entire wind farm is computed and illustrated in Fig. \ref{fig:wf}. The components of the analytical PDF are color-coded as follows: red for the probability mass at zero output power, black for the continuous region between \( P_{min} \) and \( P_{max} \), and blue for the probability mass at the maximum output power. In this analytical PDF, the probability that the output power is zero is 0.162025. Of this, 0.05 corresponds to wind power plant unavailability, and 0.112025 is due to wind speeds below the cut-in velocity ($v_{c-in}$) and above cut-off velocity ($v_{c-off}$). The continuous portion of the PDF has an area of 0.700421, representing the operating region between cut-in ($v_{c-in}$) and rated ($v_{rated}$) speeds. Finally, the discrete probability mass at \( P_{\max} \) is 0.137554, corresponding to wind speeds between the rated velocity \( v_{\text{rated}} \) and  \(v_{c-out}\). The sum of these probabilities equals to 1, as expected for a valid PDF.

To validate this model, Monte Carlo simulation is performed. Wind speed samples are drawn from the Weibull distribution, and the output power is computed based on the turbine characteristics and availability of the wind farm. The resulting empirical distribution is shown in the Fig. \ref{fig:wf} as light green histogram bars. As observed, the simulation results closely align with the analytical PDF, demonstrating that the proposed model accurately represents the wind farm’s power output behavior.

\begin{figure}[b]
\centering
\includegraphics[width=0.99\columnwidth]{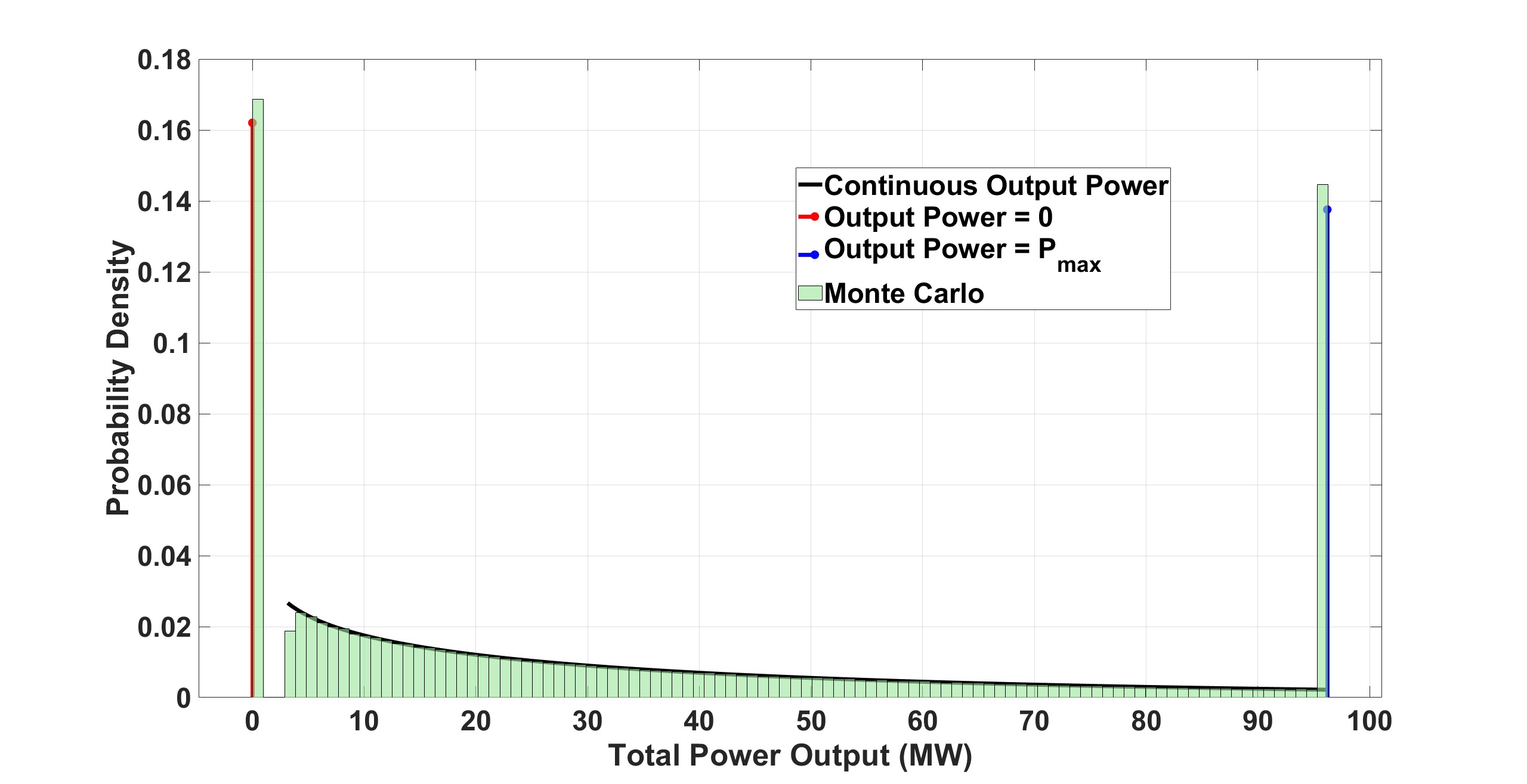}
\vspace{-0.6cm}
\caption{Analytical vs. Monte Carlo simulation of wind power PDF}
\label{fig:wf}
\vspace{-0.0cm}
\end{figure}

\item \textbf{Solar power plant}: It is modeled based on the parameters of a single solar panel, as summarized in Table \ref{tab:solar_param}. The plant is assumed to comprise 50{,}000 identical solar panels, where all panels are subject to the same solar radiation. An availability rate of 0.95 is assumed for the solar panels, accounting only for equipment failures and not the periods without sunlight during nighttime. Additionally, the solar irradiance is assumed to follow a Gamma distribution characterized by a shape parameter \( \alpha = 3.8 \) and a scale parameter \( \beta = 80 \), Berlin solar irradiance data obtained from \cite{pvgis2025}.

\begin{table}
\vspace{-0.1cm}
\centering
\caption{Solar Panel Parameters}
\label{tab:solar_param}
\begin{tabular}{ccc}
\hline
\textbf{Parameter} & \textbf{Description} & \textbf{Value} \\
\hline
\( \eta \)       & 0.2           & Conversion efficiency of a panel \\
\( \text{area} \) & 2.0 m\(^2\)   & Area of a single panel \\
\hline
\end{tabular}
\vspace{-0.2cm}
\end{table}

Using the analytical PDF derived in Equation \eqref{eq:16} and assuming that the solar panel output power does not reach its maximum value, the equivalent power output PDF of the entire solar power plant is computed and illustrated in Fig. \ref{fig:spp}. The components of the analytical PDF are represented using distinct colors: red denotes the probability mass at zero output power, while black indicates the continuous part of the distribution. 

In this analytical PDF, the probability that the output power is zero is 0.05 corresponds to solar power plant unavailability. The continuous portion of the PDF has an area of 0.95, representing the operating region. The sum of these probabilities equals 1, as expected for a valid probability density function.

To validate the analytical model, a Monte Carlo simulation is conducted. In this simulation, solar irradiance values are randomly sampled from the assumed Gamma distribution, and the corresponding power output is calculated based on the characteristics of the solar panels and the availability of the plant. The resulting empirical distribution is displayed in Fig. \ref{fig:spp} as yellow histogram bars. As evident, the simulated results exhibit strong agreement with the analytical PDF, confirming that the proposed model provides an accurate representation of the power output behavior of the solar power plant.

\begin{figure}
\centering
\includegraphics[width=0.99\columnwidth]{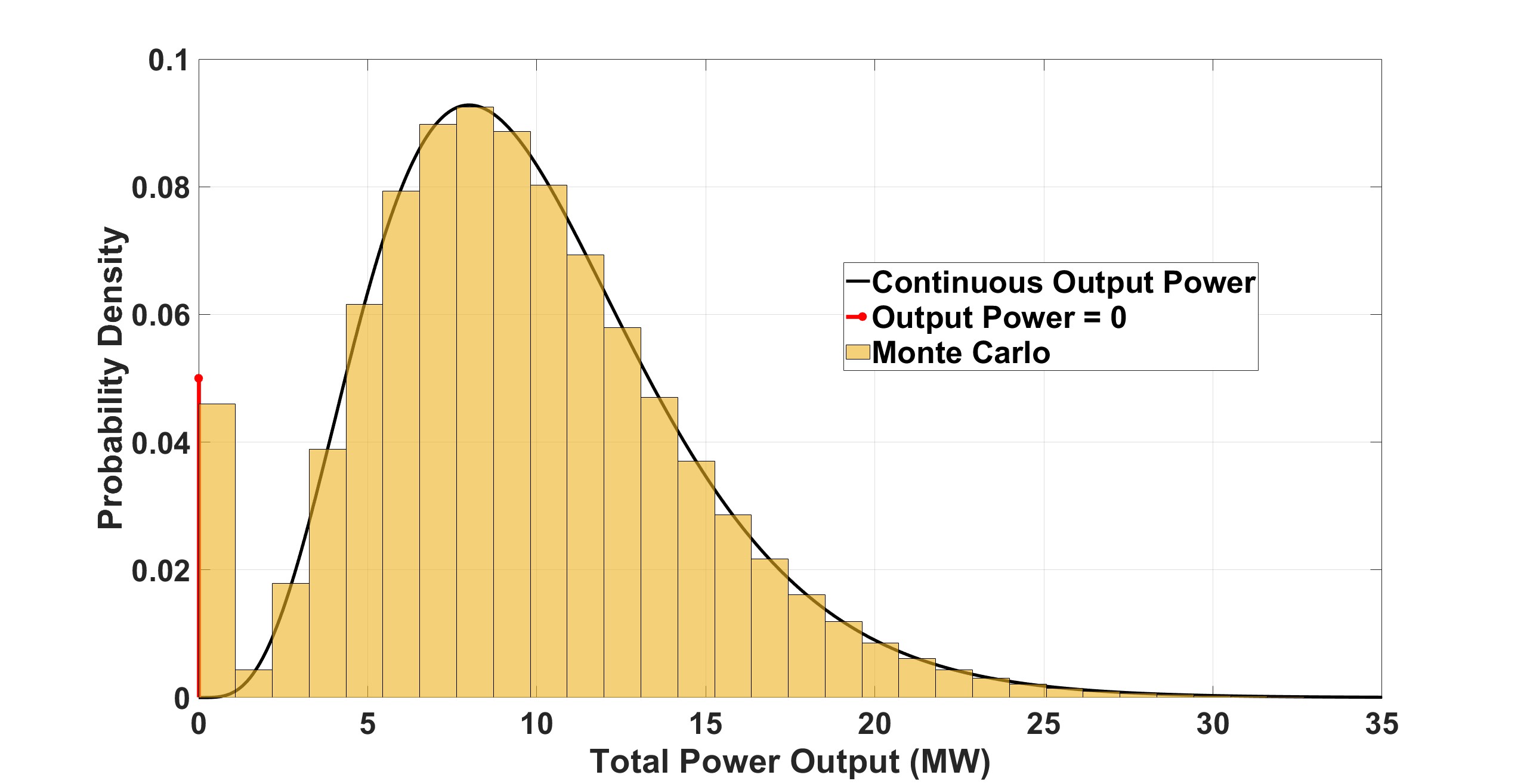}
\vspace{-0.7cm}
\caption{Analytical PDF vs. Monte Carlo simulation of solar power output}
\label{fig:spp}
\vspace{-0.02cm}
\end{figure}

\end{itemize}

\vspace{-0.35cm}
\subsection{Validation of Proposed Methods for the Entire Power System}
\label{ssec:s2}
As the share of renewable energy continues to grow in real-world power systems, the IEEE RTS-96 has been modified to reflect this trend. Specifically, several conventional power plants have been replaced by wind and solar power plants, as detailed in Table \ref{tab:ieee_new}. It should be noted that conventional power plants are replaced by renewable sources in such a way that the expected total power output remains approximately the same. This ensures that the LOLE stays comparable between the original system with conventional plants and the modified system with renewables. In this modified system, renewable generation accounts for approximately 47\% of the total installed generation capacity. In this table, the gray, green, and yellow shaded cells represent conventional, wind, and solar power plants, respectively. In this study, the availability of wind power plants is assumed to be 0.95. For solar power plants, accounting for both panel performance and daylight availability, the availability is assumed to be 0.475.

\begin{table}[b]
\vspace{-0.1cm}
\centering
\caption{Modified Generating Unit Locations}
\label{tab:ieee_new}
\resizebox{\columnwidth}{!}{%
\begin{tabular}{ccccccc}
\hline
Bus & Unit 1  & Unit 2  & Unit 3  & Unit 4  & Unit 5  & Unit 6  \\ 
    & (MW)    &  (MW)   &  (MW)   &  (MW)   &  (MW)   &  (MW)   \\ \hline
1   & \cellcolor{yellow!40}20 (S1) & \cellcolor{gray!40}20 & \cellcolor{gray!40}76 & \cellcolor{green!40}76 (W1)& & \\ 
2   & \cellcolor{yellow!40}20 (S2) & \cellcolor{gray!40}20 & \cellcolor{gray!40}76 & \cellcolor{green!40}76 (W2)& & \\ 
7   & \cellcolor{gray!40}100 & \cellcolor{green!40}100 (W3)& \cellcolor{yellow!40}100 (S3) & & & \\ 
13  & \cellcolor{gray!40}197 & \cellcolor{green!40}197 (W4)& \cellcolor{green!40}197 (W5)& & & \\ 
15  & \cellcolor{gray!40}12 & \cellcolor{gray!40}12 & \cellcolor{gray!40}12 & \cellcolor{gray!40}12 & \cellcolor{gray!40}12  & \cellcolor{green!40}155 (W6)\\ 
16  & \cellcolor{green!40}155 (W7)& & & & & \\ 
18  & \cellcolor{gray!40}400 & & & & & \\ 
21  & \cellcolor{gray!40}400 & & & & & \\ 
22  & \cellcolor{green!40}50 (W8)& \cellcolor{green!40}50 (W9)& \cellcolor{green!40}50 (W10)& \cellcolor{gray!40}50 & \cellcolor{gray!40}50 & \cellcolor{yellow!40}50 (S4) \\ 
23  & \cellcolor{green!40}155 (W11)& \cellcolor{yellow!40}155 (S5)& \cellcolor{gray!40}350 & & & \\ \hline
\end{tabular}%
}
\vspace{-0.2cm}
\end{table}

For renewable power plants, the codes in parentheses indicate the wind speed or solar irradiance dataset associated with each plant. For the wind power plants, wind speed at each location is modeled using a Weibull distribution, with the scale and shape parameters provided in Table \ref{tab:wind}. Similarly, solar irradiance at solar plant locations is modeled using a Gamma distribution, with the corresponding distribution parameters listed in Table \ref{tab:solar}.

\begin{table}
\vspace{-0.1cm}
\centering
\caption{Weibull wind speed parameters by location \cite{nasa}}
\label{tab:wind}
\resizebox{\columnwidth}{!}{
\begin{tabular}{cccc}
\hline
Wind Unit & Scale Parameter ($c$) & Shape Parameter ($k$) & Location\\
\hline
W1  & 7.2  & 2.0 & Esbjerg (Denamrk)\\
W2  & 7.8  & 2.1 & Texas (USA)\\
W3  & 8.0  & 1.8 & Moray Firth (Scotland)\\
W4  & 8.0  & 2.1 & Patagonia (Argentina)\\
W5  & 7.8  & 2.2 & Inner Mongolia (China)\\
W6  & 9.0  & 2.5 & North Sea (UK)\\
W7  & 8.8  & 2.1 & California (USA)\\
W8  & 6.5  & 1.9 & Andros (Greece)\\
W9  & 7.0  & 2.1 & Wladyslawowo (Poland)\\
W10 & 6.8  & 2.0 & Oklahoma (USA)\\
W11 & 9.1  & 2.4 & Tasmania (Australia)\\
\hline
\end{tabular}
}
\vspace{-0.02cm}
\end{table}

\begin{table}[b]
\vspace{-0.1cm}
\centering
\caption{Gamma solar irradiance parameters by location \cite{pvgis2025}}
\label{tab:solar}
\begin{tabular}{cccc}
\hline
Solar Unit & Alpha ($\alpha$) & Beta ($\beta$) & Location \\
\hline
S1 & 3.8 & 80 & Berlin (Germany) \\
S2 & 3.6 & 82 & London (UK)\\
S3 & 4.2 & 85 & Rome (Italy) \\
S4 & 2.9 & 78 & Copenhagen (Denmark)\\
S5 & 4.8 & 70 & New York (USA)\\
\hline
\end{tabular}
\vspace{-0.3cm}
\end{table}

In this part, the LOLE and EENS of the system are calculated using three different states as follows.

State I applies the central limit theorem to model the total system output power as an exact normal distribution. State II refines this approximation by using the Gram–Charlier expansion, as detailed in Subsection \ref{ssec:Gram-Charlier series}, to account for deviations from normality. In both States I and II, stress–strength interference theory is employed to calculate the LOLE and EENS. Finally, state III uses the Monte Carlo simulation to validate the accuracy of the previous methods. A relative error of 0.5\% was used as the convergence criterion to ensure statistically accurate estimates.

\begin{figure}
\centering
\includegraphics[width=1.0\columnwidth]{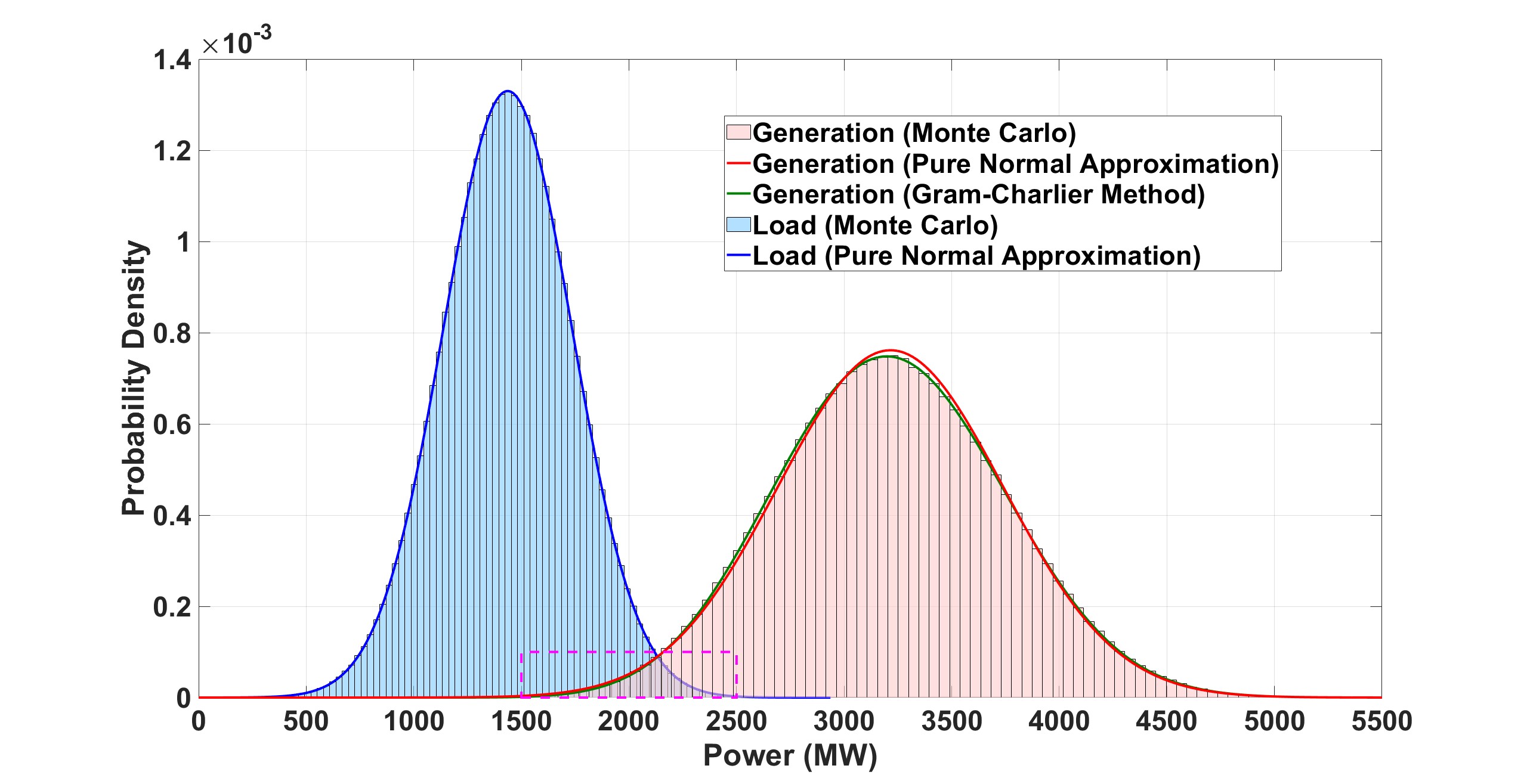}
\vspace{-0.8cm}
\caption{Load and generation distributions fitted by proposed pure normal and Gram–Charlier methods}
\label{fig:S2}
\vspace{-0.1cm}
\end{figure}

In this way, Fig.~\ref{fig:S2} shows the histograms of the system load and generation, represented by light blue and pink bars, respectively, obtained from Monte Carlo simulation. A normal distribution curve has been fitted to the load histogram and is shown as a blue curve. For the generation histogram, two curves are fitted: the red curve assumes that the generation PDF follows a pure normal distribution, while the green curve accounts for deviations from normality using the Gram–Charlier approximation.

\begin{figure}[t]
\centering
\includegraphics[width=1.0\columnwidth]{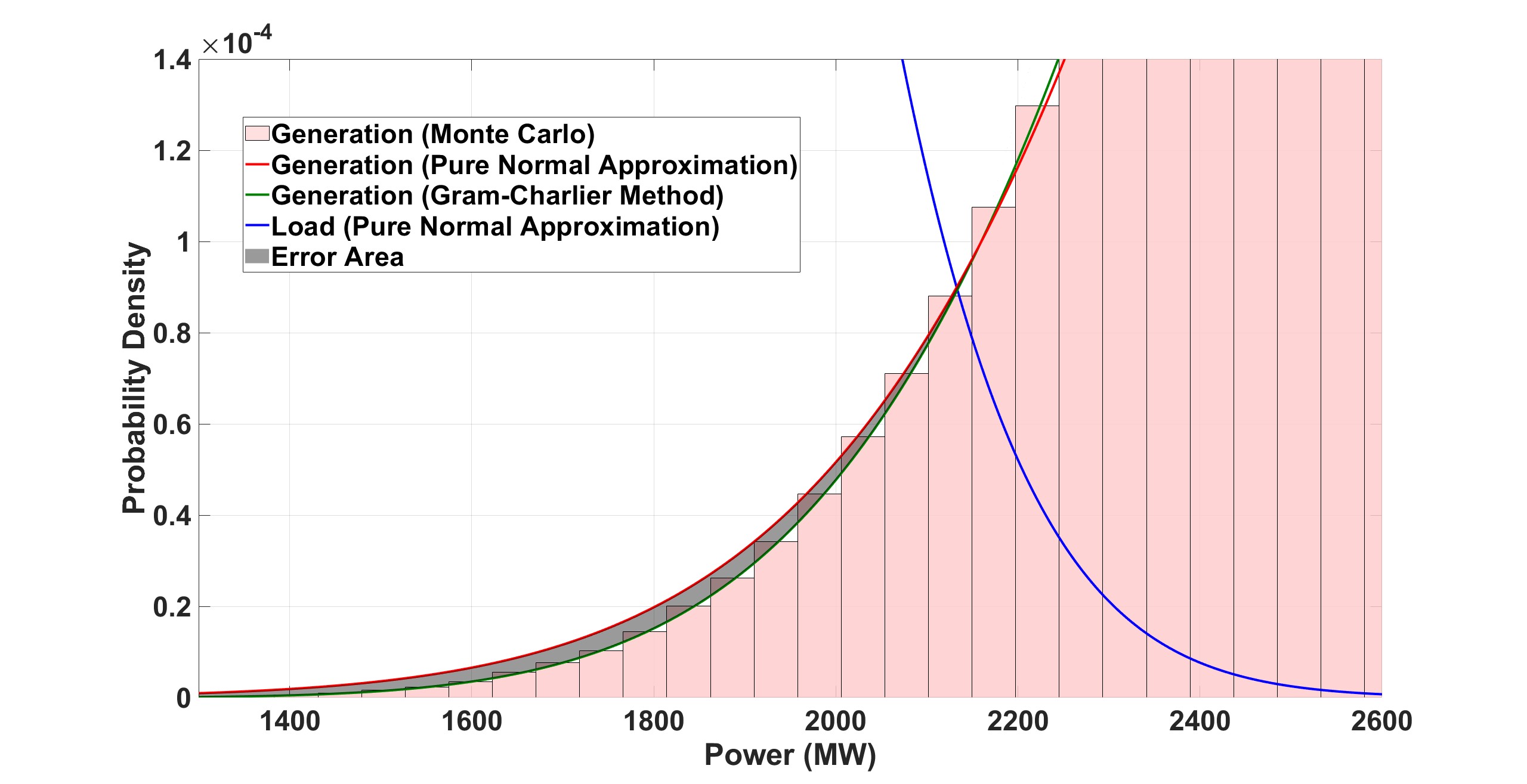}
\vspace{-0.7cm}
\caption{Zoomed in view of generation/load distribution from Fig. \ref{fig:S2}}
\label{fig:S2_1}
\vspace{-0.1cm}
\end{figure}

The area highlighted by the magenta rectangle in Fig. \ref{fig:S2} has been zoomed in Fig. \ref{fig:S2_1} for better clarity. In this figure, the histogram of the load has been removed, while the normal fit curve is retained. As shown, the green curve—representing the proposed generation model, passes through approximately the center of the bars in the generation histogram. This suggests that the results obtained from the proposed method closely match those from the Monte Carlo simulation. The gray shaded area in Fig. \ref{fig:S2_1} illustrates the difference between the normal fit curve of the generation and the curve from the proposed method, Gram-Charlier expansion.

Table \ref{tab:comparison} presents a comparative analysis of the running time and system reliability metrics, LOLE and EENS, calculated using three different methods: Monte Carlo simulation, the proposed Gram–Charlier–based method, and the central limit theorem approach, in which the obtained distribution is assumed to be purely normal. All simulations and computations were performed on a high-performance workstation equipped with dual Intel\textsuperscript{\textregistered} Xeon\textsuperscript{\textregistered} Silver 4210R CPUs (2.40\,GHz), 192\,GB of RAM, and a 64-bit Windows operating system. Under these conditions, the LOLE and EENS for the modified IEEE RTS-96 test systems were calculated as 9.3868~(hrs/year) and 1362.12~(MWh), respectively, using Monte Carlo simulation.  The simulation ran for 104 seconds. These Monte Carlo results are used as the reference baseline for subsequent comparisons.

Then, the proposed method achieves accurate estimates for both LOLE and EENS, with minimal relative errors of 2.37\% and 4.27\%, respectively. This demonstrates that the method maintains high reliability assessment precision. Additionally, the computational efficiency is significantly improved, reducing the run-time from over 104 seconds, using traditional Monte Carlo simulation, to just 1.1 seconds, making it highly suitable for large-scale or real-time applications. In contrast, the pure normal assumption, while computationally efficient, results in substantial inaccuracies—yielding relative errors of 49.75\% for LOLE and 78.72\% for EENS. These findings highlight the limitations of treating the distribution as exactly normal when representing the stochastic nature of power generation, particularly under the high variability and elevated failure rates associated with renewable integration.

A detailed comparison of the LOLE and EENS results, along with their corresponding relative errors and computation times, is summarized in Table~\ref{tab:comparison}.

\begin{table}
\vspace{-0.03cm}
\centering
\caption{Comparison of LOLE, EENS, and running time across methods (Renewable penetration = 47 \%)}
\label{tab:comparison}
\resizebox{\columnwidth}{!}{
\begin{tabular}{lccc}
\toprule
Indices & \begin{tabular}[c]{@{}c@{}}Monte Carlo\\Simulation\end{tabular} & \begin{tabular}[c]{@{}c@{}}Gram--Charlier\\Method\end{tabular} & \begin{tabular}[c]{@{}c@{}}Pure Normal\\Approximation\end{tabular} \\
\midrule
LOLE (hrs/yr)       & 9.3868  & 9.1647  & 14.057 \\
LOLE Error (\%)     & 0       & 2.37    & 49.75 \\
EENS (MWhr)         & 1362.12 & 1303.93 & 2434.44 \\
EENS Error (\%)     & 0       & 4.27    & 78.72 \\
Running Time (s)    & 104.11  & 1.01    & 0.38 \\
\bottomrule
\end{tabular}
}
\vspace{-0.2cm}
\end{table}

\vspace{-0.4cm}
\subsection{Reliability Assessment under Increasing Renewable Integration}
\label{ssec:s3}

In this part, the impact of increasing the share of renewable power plant is investigated. As a first step, the 100 MW unit at bus 7 and the 400 MW unit at bus 18 are replaced by wind power plants W3 and W6, respectively, as defined in Table \ref{tab:wind}. This substitution raises the total renewable share to {61.8\%}. The resulting LOLE and EENS values, calculated using the three methods described earlier, are presented in Table \ref{tab:comparison2}. As illustrated, increasing the share of renewable energy sources reduces the estimation error for both the proposed Gram–Charlier–based method and the central limit theorem approach, in which the obtained distribution is assumed to be purely normal. Moreover, the run-time of the Monte Carlo simulation increases significantly, by approximately 14\%, as renewable penetration grows. In contrast, the proposed method exhibits only a modest increase in computation time of about 2\%, demonstrating its scalability and computational efficiency.

\begin{table}[b]
\vspace{-0.1cm}
\centering
\caption{Comparison of LOLE, EENS, and running time across methods (Renewable penetration = 61.8 \%)}
\label{tab:comparison2}
\resizebox{\columnwidth}{!}{
\begin{tabular}{lccc}
\toprule
Indices & \begin{tabular}[c]{@{}c@{}}Monte Carlo\\Simulation\end{tabular} & \begin{tabular}[c]{@{}c@{}}Gram--Charlier\\Method\end{tabular} & \begin{tabular}[c]{@{}c@{}}Pure Normal\\Approximation\end{tabular} \\
\midrule
LOLE (hrs/yr)       & 9.3868  & 9.172  & 13.878 \\
LOLE Error (\%)     & 0       & 2.28  & 47.85 \\
EENS (MWhr)         & 1362.12 & 1306.01 & 2393.79 \\
EENS Error (\%)     & 0       & 4.12  & 75.74 \\
Running Time (s)    & 119.13  & 1.03  & 0.38 \\
\bottomrule
\end{tabular}
}
\vspace{-0.2cm}
\end{table}

\begin{figure}
\centering
\includegraphics[width=.9\columnwidth]{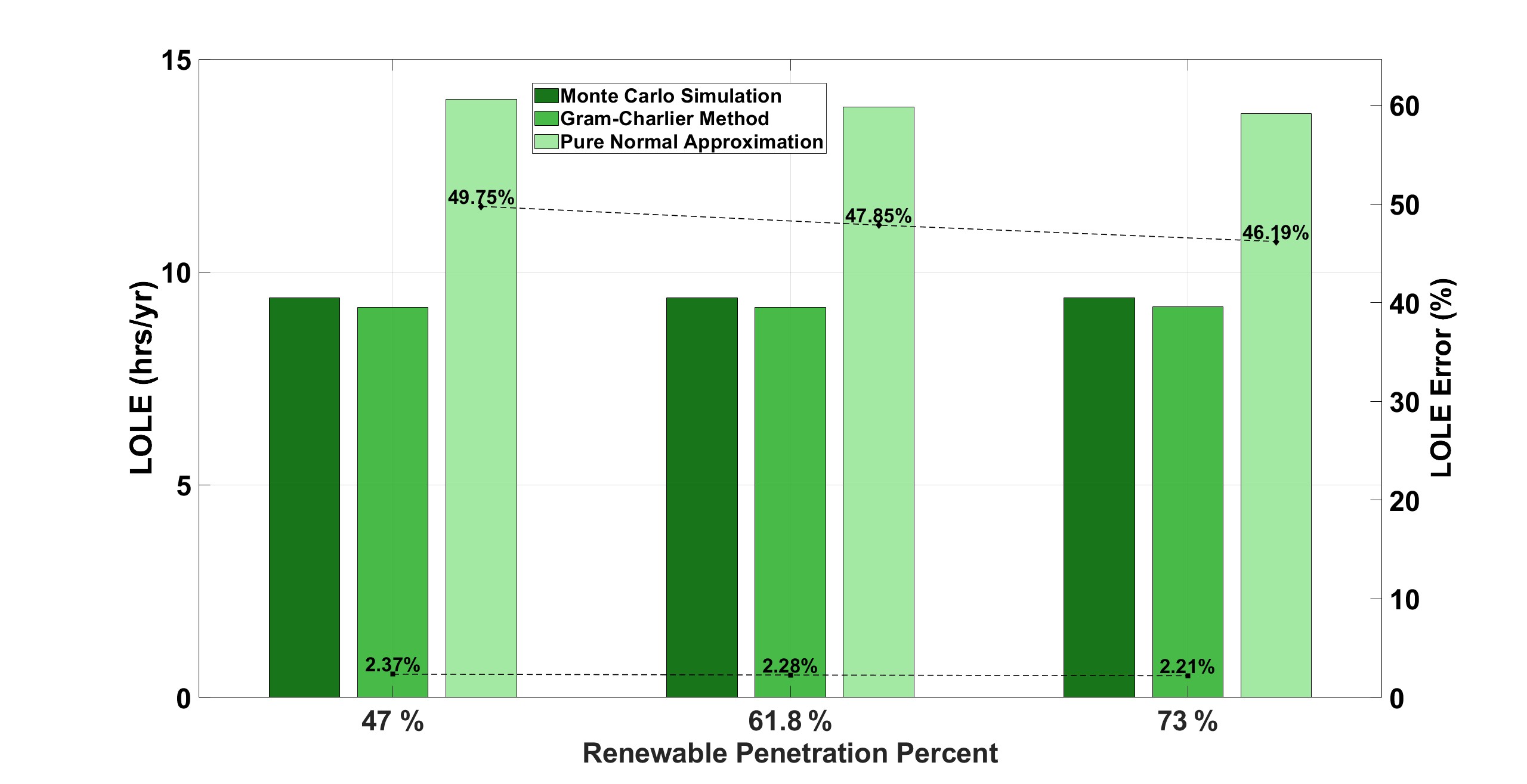}
\vspace{-0.06cm}
\caption{LOLE and LOLE error trends with increasing renewable share}
\label{fig:comparison}
\vspace{-0.01cm}
\end{figure}

\begin{figure}
\vspace{-0.03cm}
\centering
\includegraphics[width=.9\columnwidth]{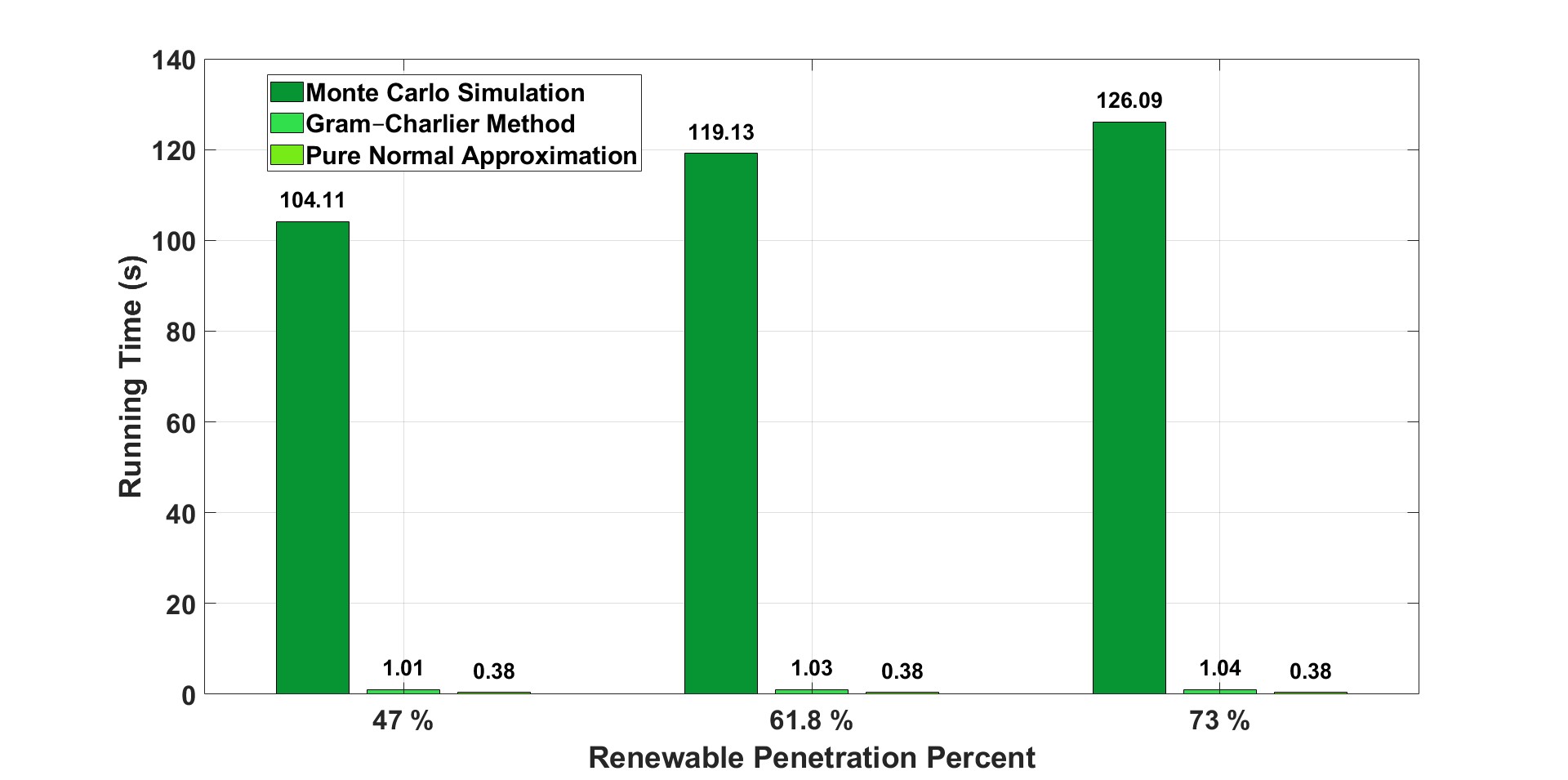}
\vspace{-0.06cm}
\caption{Running time trends with increasing renewable share}
\label{fig:comparison2}
\vspace{-0.1cm}
\end{figure}

In the second step, the renewable share is further increased by replacing the 400 MW unit at bus 21 with a wind power plant (W6 in Table \ref{tab:wind}). As a result, the total share of renewables rises to 73\%. Similarly to the previous case, the values and errors for LOLE and EENS decrease, as shown in Table \ref{tab:comparison3}. Fig. \ref{fig:comparison} and Fig. \ref{fig:comparison2} summarize the outcomes across different levels of renewable penetration. As observed, increasing the share of renewable energy leads to reduced estimation errors for LOLE  when using the Gram–Charlier–based method and the pure normal approximation. In contrast, the computational time of the Monte Carlo simulation increases significantly with higher levels of renewable integration. The proposed method, however, maintains high efficiency, exhibiting only a marginal increase in execution time. In the context of the ongoing global transition toward renewable energy in modern power systems, the proposed approach demonstrates superior accuracy and computational scalability under high renewable penetration scenarios.

\begin{table}
\vspace{-0.03cm}
\centering
\caption{Comparison of LOLE, EENS, and running time across methods (Renewable penetration = 73 \%)}
\label{tab:comparison3}
\resizebox{\columnwidth}{!}{
\begin{tabular}{lccc}
\toprule
Indices & \begin{tabular}[c]{@{}c@{}}Monte Carlo\\Simulation\end{tabular} & \begin{tabular}[c]{@{}c@{}}Gram--Charlier\\Method\end{tabular} & \begin{tabular}[c]{@{}c@{}}Pure Normal\\Approximation\end{tabular} \\
\midrule
LOLE (hrs/yr)       & 9.3868  & 9.179  & 13.722 \\
LOLE Error (\%)     & 0       & 2.21  & 46.19 \\
EENS (MWhr)         & 1362.12 & 1308.18 & 2357.69 \\
EENS Error (\%)     & 0       & 3.96  & 73.09 \\
Running Time (s)    & 126.09 & 1.04  & 0.38 \\
\bottomrule
\end{tabular}
}
\vspace{-0.2cm}
\end{table}

\begin{table*}[t]
\centering
\caption{Accuracy and Computational Performance Comparison for Different Systems (Renewable Penetration = 61.8 \%)}
\label{tab:reliability_comparison}
\begin{tabular}{ccccccc}
\toprule
& \multicolumn{3}{c}{Monte Carlo} 
& \multicolumn{3}{c}{Gram--Charlier}  \\
\cmidrule(lr){2-4} \cmidrule(lr){5-7}
System 
& LOLE (hrs/yr)  & EENS (MWhr)  & Running Time (s)
& LOLE (hrs/yr)  & EENS (MWhr)  & Running Time (s)
 \\
\midrule
IEEE 1-Area RTS-96           &9.3868  &1362.12  &119.13  &9.172  &1306.01  &1.03  \\
IEEE 3-Area RTS-96     &9.3374  &4085.67  &381.71  &9.147  &3918.68  &2.83  \\
IEEE 5-Area RTS-96     &9.3513  &6811.14  &635.09  &9.156  &6530.04  &4.75  \\
\bottomrule
\end{tabular}
\end{table*}

\subsection{Evaluation of Accuracy and Scalability of the Proposed Method}
\label{ssec:s4}
To comprehensively evaluate the accuracy and scalability of the proposed method, this section examines the IEEE three-area RTS-96 and five-area RTS-96 systems, which are three and five times larger than the original IEEE one area RTS-96, respectively \cite{780914}. The characteristics of the generation units and the associated renewable energy parameters for the IEEE one area RTS-96 system are detailed in Tables \ref{tab:ieee_new} and \ref{tab:wind}, providing a foundation for the extended multi-area systems. In this study, the share of renewable generation is assumed to be 61.8\% of the total installed capacity. Table \ref{tab:reliability_comparison} summarizes the performance results in terms of LOLE (hrs/yr), EENS error (MWhr), and computational running time (s). The results clearly show that the computational time increases significantly with system size when the Monte Carlo simulation method is used. This behavior is expected due to the large number of samples required to obtain accurate reliability estimates in larger systems.

In contrast, the proposed methodology delivers accurate reliability results while efficiently handling the increased complexity of larger systems. These findings demonstrate its robustness and scalability, confirming its suitability for reliability assessment of large and complex power system models without compromising accuracy or computational efficiency.

\vspace{-0.04cm}
\section{Conclusion}
\label{sec:Conclusion}

This paper presented an efficient method for resource adequacy assessment in renewable-rich power systems. The proposed framework models the probabilistic behavior of conventional and renewable power plants using their respective probability distribution functions, from which statistical moments and cumulants are derived. The Gram–Charlier expansion is then used to better represent the generation distribution, particularly in situations where the normal approximation becomes inaccurate due to the limited number of generation units and the high failure rates associated with renewable energy resources. Reliability indices, including LOLE and EENS, are evaluated using the stress–strength method. The proposed approach was validated using the modified IEEE one area RTS-96 and its multi-area configurations. The results demonstrate that the method can accurately estimate adequacy indices while maintaining significantly lower computational time compared with Monte Carlo simulation. In the studied scenarios, the calculated LOLE and EENS values showed strong agreement with Monte Carlo simulation results, while the conventional normal approximation produced substantially larger errors. These results underscore the importance of improved probabilistic representations for systems with high renewable penetration. The proposed method remains effective as renewable penetration level increase and scales well to larger systems, making it suitable for future power system studies and operational planning. Future work will focus on incorporating transmission and distribution constraints and exploring data-driven approaches that estimate generation distributions directly from empirical data to further improve modeling accuracy.
\vspace{-0.03cm}
\printbibliography

\end{document}